\documentclass[12pt]{article}
\usepackage{comment,amsmath,graphpap,amsthm,amssymb,latexsym,epsfig}
\usepackage{finalT}

\numberwithin{equation}{section}

\input newtonianlimit.def

\begin{document}

\title{Existence of families of spacetimes  with a Newtonian
limit}

\author{Todd A. Oliynyk \thanks{Todd.Oliynyk@sci.monash.edu.au}\\
School of Mathematical Sciences\\
Monash University VIC 3800\\
Australia
\and
Bernd Schmidt \thanks{Bernd.Schmidt@aei.mpg.de}\\
Max Planck Institute \\
for Gravitational Physics \\
Am M\"uhlenberg 1\\
D-14476 Golm\\
Germany}

\date{}
\maketitle

\begin{abstract}
J\"urgen Ehlers developed \emph{frame theory} to better understand the
relationship between general relativity and Newtonian gravity. Frame theory contains a parameter $\lambda$, which can be thought
of as $1/c^2$, where $c$ is the speed of light. By construction, frame theory is equivalent to general relativity for $\lambda >0$,
and reduces to Newtonian gravity for $\lambda =0$. Moreover, by setting  $\ep=\sqrt{\lambda}$, frame theory provides a
framework to study the Newtonian limit $\ep \searrow 0$  (i.e. $c\rightarrow \infty$). A number of ideas
relating to frame theory that were introduced by J\"urgen have subsequently found important applications to the rigorous
study of both the Newtonian limit and post-Newtonian expansions. In this article, we review frame theory and discuss, in
a non-technical fashion, some of the rigorous results on the Newtonian limit and post-Newtonian expansions that have
followed from J\"urgen's work.
\end{abstract}

\section{Introduction}
Throughout his scientific career, J\"urgen was interested in conceptual questions of physics, and in particular, the relationship between different theories
that describe the same phenomena in nature. A prime example of this is the solar system. It can be, to high accuracy, described using either Newton's or Einstein's theory of gravity.
The question then becomes how can we understand this relationship between the two theories in spite of their quite distinct formulations.

To begin to answer this question, J\"urgen  developed, extending ideas of Cartan \cite{ca} and Friedrichs  \cite{fr}, a theory, called \emph{frame theory}, which contains both Einstein's and Newton's theories of gravity. Frame theory contains a parameter $\lambda$, which can be thought of as $1/c^2$ where $c$ is the speed of light. By construction, frame theory  coincides with Einstein's theory for $\lambda >0$, and reduces to Newton's theory when
$\lambda=0$. In this way, frame theory provides a way of understanding the relationship between the \emph{laws}, or, in other words, the equations of the two theories.

\begin{figure}
\begin{center}
\epsfig{figure=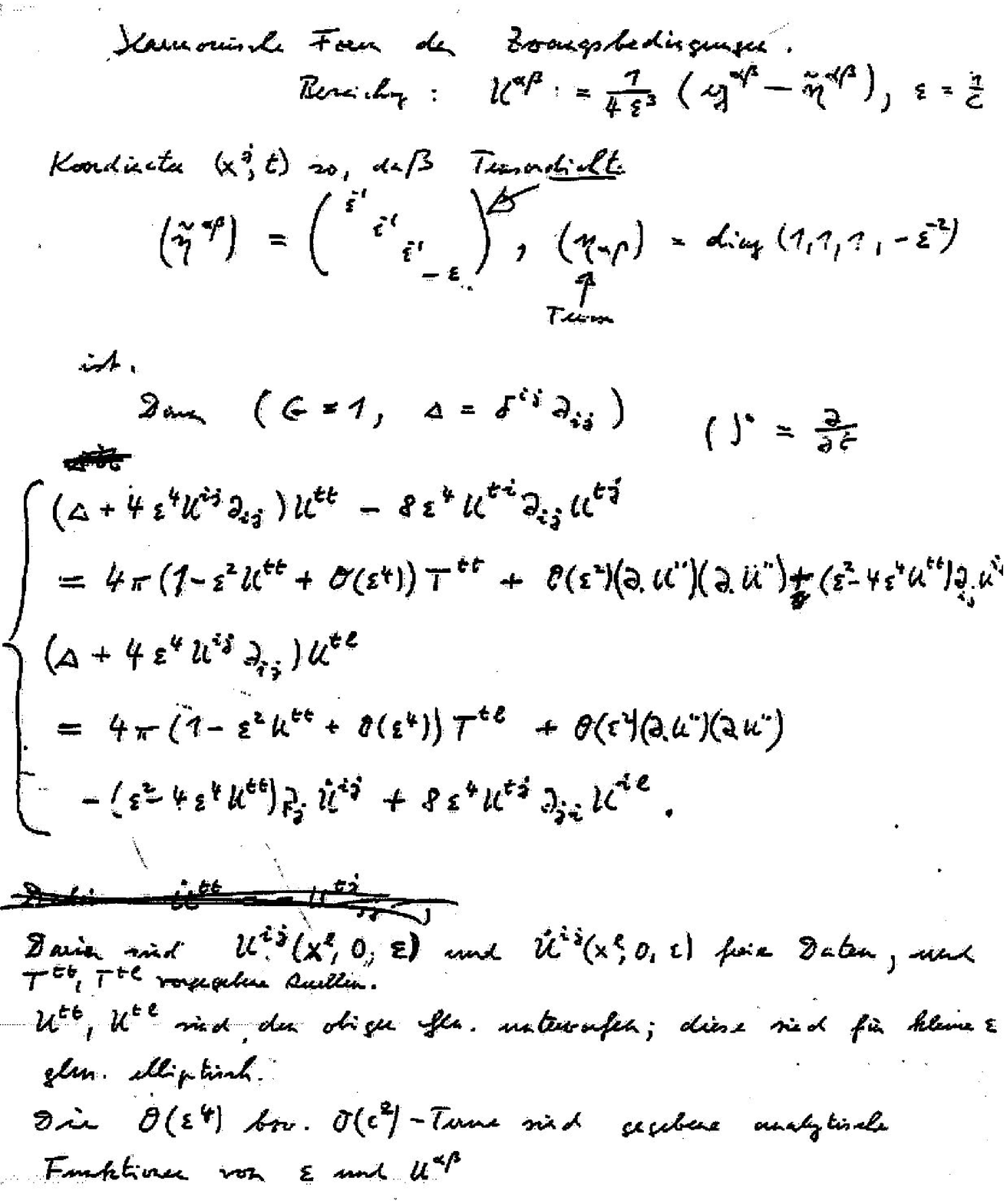,height=17.0cm}

\caption{J\"urgen's notes where he introduces the gravitational variables \eqref{EhlersvarsA}-\eqref{EhlersvarsB}.} \label{JEnotes}
\end{center}
\end{figure}

In addition to developing frame theory,  J\"urgen made another important contribution which he never published. He discovered
(see Figure \ref{JEnotes} at the end of the article)  a way of writing
Einstein's field equations containing the parameter
$\epsilon =\sqrt{\lambda}$
for which the partial differential equations of Einstein's theory ($\epsilon>0$) become the equations of Newton's theory in the
limit $\epsilon\searrow 0$. Following standard terminology,
we will refer to the limit $\ep \searrow 0$ as the \emph{Newtonian limit}. 

Although, J\"urgen's formulation provided a satisfying method for addressing the relationship between the two theories at the level of equations, the problem of
understanding the relationship between the solutions remained open. The relationship between the solutions is what one really wants to know
as it is the solutions that describe the physical phenomena, and not the equations themselves.
A first result in this direction was obtained by Lottermoser \cite{Lott}. Using J\"urgen's formulation, Lottermoser showed that there exists one-parameter families of solutions
depending on $\ep$ to the gravitational constraint equations that converge in the limit $\ep \searrow 0$ to Newtonian initial data. This result was a necessary
first step in the analysis of the limit $\ep \searrow 0$ since it established the existence of reasonable, relativistic initial data that depends on $\ep$ in the proper fashion.
However, Lottermoser did not answer the existence question for relativistic solutions with a Newtonian limit. The first general result showing the existence
of a Newtonian limit for any matter model was by
Rendall for Vlasov matter. In \cite{Ren94}, Rendall rigorously proved that Einstein-Vlasov system
has a large class of solutions that have a well defined Newtonian limit.

Another important existence result for the Newtonian limit, this time in the stationary setting, was obtained by Heilig \cite{Heil}. Heilig proved the existence of stationary, rotating,
axi-symmetric solutions of the Einstein-Euler equations using J\"urgen's formulation to perturb away from  a Newtonian solution at $\ep = 0$ to obtain a fully relativistic solution
for $\ep>0$. This
paper still represents the only known existence proof for stationary, fully relativistic, perfect fluid solutions. Heilig also established that these solutions depend analytically on
$\ep$, and therefore, admit convergent post-Newtonian expansions. Later, Heilig's methods were adapted by one of the authors to prove existence of static
solutions for the Einstein field equations
coupled to Yang-Mills fields \cite{Oli05a,Oli05b}.

In \cite{Oli06}, one of the authors adapted the variables introduced by J\"urgen to prove the existence of a wide class of solutions to the Einstein-Euler equations
that have a Newtonian limit. We note that this formulation was different from Rendall's both in terms of the variable used to represent the gravitational
field, and the choice of gauge.
The methods used in \cite{Oli06} were subsequently generalized to prove the existence of a large class of
solutions to the Einstein-Euler equations that admit a first post-Newtonian expansion \cite{Oli07}. The techniques used in \cite{Oli06} and \cite{Oli07} are
based on  energy estimates derived from a first order nonlocal formulation
of the Einstein-Euler equations in the harmonic gauge. A further improvement was achieved in \cite{Oli08} through the use
of dispersive wave estimates in addition to the energy estimates. This improvement allowed the conditions on the
initial data need to ensure the existence of a Newtonian limit to be relaxed as compared to \cite{Oli06}.
It is also possible to use the dispersive wave estimates to extend the results of
\cite{Oli08}, and establish the existence of post-Newtonian expansions to the second order.

The main aim on this paper is to describe, in a non-technical fashion, the rigorous
results obtained in \cite{Oli06,Oli07,Oli08,Oli09} on the Newtonian limit and post-Newtonian expansions, and moreover, to relate these
results to frame theory. These rigorous results, which originate from ideas introduced by J\"urgen, show that the question about the relationship between
Einstein's and Newton's theories of gravity is not just philosophical, but in fact, leads to a deeper insight into the structure of the equations, and a better
understanding of both the Newtonian limit and
post-Newtonian approximations. We devote this article to J\"urgen's memory.

\section{The Frame Theory}

In this section, we briefly describe frame theory closely following J\"urgen's paper \cite{Ehl97}.
Frame theory is formulated using the following collections of fields on a \emph{spacetime} (i.e. a $4$--manifold):
\begin{itemize}
\item[] $t_{ik}$ a nowhere vanishing symmetric, $2$-covariant tensor field,

\item[]  $s^{ik}$ a nowhere vanishing symmetric, $2$-contravariant tensor field,

\item[]  $\Gamma^l_{ik}$ a symmetric linear connection, and

\item[] $T^{ik}$ a  symmetric, $2$-contravariant tensor field.

\end{itemize}
The fields $t_{ik}$, $s^{ik}$, $\Gamma^l_{ik}$, and $T^{ik}$ are referred to as the
the \emph{temporal metric},  \emph{inverse temporal metric}, \emph{gravitational field}, and \emph{matter tensor}, respectively.
The laws of frame theory contain these fields and two real-valued parameters $\lambda$ and $\Lambda$. To formulate the laws, it is convenient ``raise'' and
``lower'' tensorial indices using $s^{ik}$ and $t_{ik}$, respectively. For example, we define
\begin{equation*}
V^\bullet_k:=t_{ki}V^i, \AND \omega_\bullet^{ik}:=s^{ik}\omega_k.
\end{equation*}
To avoid confusion it is necessary to indicate by dots whether the original object has upper or lower indices, as shown.

\bigskip

\noindent The (local) laws of frame theory are:

\begin{itemize}

\item[(i)] For any timelike vector $V^i$ (i.e. $t_{ik}V^iV^k>0$), the quadratic form
$\omega_i\to s^{ik}\omega_i\omega_k$ is positive definite on the subspace $\{\, \omega_i\,| \,\omega_iV^i=0\, \}$ of the covector space.
\item[(ii)] The temporal and inverse temporal metrics are related to each other by
\eqn{tempmet1}{
t_{ik}s^{kl}=-\lambda\delta^l_i,
}
and satisfy
\eqn{tempmet2}{
t_{ik;l}=0, \AND \; s^{ik}{}_{;l}=0,
}
where the semicolon denotes the covariant derivative determined by the connection $\Gamma^l_{ik}$.
\item[(iii)] The curvature tensor $R^j{}_{ikj}$ of
the connection $\Gamma^l_{ik}$ satisfies
\eqn{Curve}{
R^i{}_k{}_\bullet^l{}_m=R^l{}_m{}_\bullet^i{}_k,
}
while the Ricci tensor $R_{ik}=R^j{}_{ikj}$ and the matter tensor are related by
\eqn{Ricci}{
R_{ik}=8\pi(T^{\bullet\bullet}_{ik}-{1\over 2}t_{ik}T^j{}^\bullet_j)-\Lambda t_{ik}.
}
\item[(iv)] The matter tensor satisfies
\begin{equation*}
T^{ik}{}_{;k}=0.
\end{equation*}
\item[(v)]
In addition to general laws (i)-(iv), a {\it  matter model} must be specified.  Here, we consider only the case
of a {\it perfect fluid} where
\begin{equation*}
T^{ik}=(\rho+\lambda p)U^iU^k+ps^{ik}
\end{equation*}
with  $\ t_{ik}U^iU^k=1$, $\rho>0$, and $\rho+\lambda p>0$.
\end{itemize}

\bigskip

The following facts can be inferred from  the references [3...9] in \cite{Ehl97}.
If $\lambda>0$, then, by appropriate rescaling, one can choose $\lambda=1, g_{ik}=-t_{ik},\ g^{ik}=s^{ik}$, and in this case
frame theory reduces to general relativity with the natural units $G=c=1$, and a spacelike signature $(+++-)$. For the purpose of this paper, however, it is preferable not to fix the scaling  and work with $t_{ik}=-\lambda g_{ik}, s^{ik}=g^{ik}$, and arbitrary $\lambda>0$.

If $\lambda=0$, then frame theory reduces to  Newton-Cartan theory. In this case, the temporal metric can be expressed in terms of a scalar {\it absolute time $t$} by
\eqn{Ncmet}{
t_{ik}=t_{,i}t_{,k},
}
and it is possible to choose (local) coordinates $(x^K,x^4=t)$ $K=1,2,3$ such that
\eqn{NC}{
t_{ik} =\text{\rm diag}(0,0,0,1), \quad
s^{ik} =\text{\rm diag}(1,1,1,0),
}
and the nonzero components of the connection $\Gamma^i_{kl}$ are given by
\eqn{NCchrist}{
\Gamma^K_{tt}=-g^K  \AND \Gamma^K_{tL}= s^{KF}\epsilon_{LFJ}\omega^J.
}
Here, the spatial vector fields $\vec g(\vec x,t) = (g^K(\vec x,t))$ and $\vec \omega(\vec x,t) = (\omega^K(\vec x,t))$ play the part of
the gravitational acceleration and the Coriolis angular velocity, respectively.
While Newton-Cartan theory is more general than Newtonian gravity, it does reduces to Newton's theory if $\vec\omega$ only depends
on $t$. This is due to the existence
of  non-rotating coordinates for which $\vec\omega=0$, and
\begin{equation*}
\Gamma^i_{kl}=t_{,k}t_{,l}s^{ij}\Phi_{,j}\ ,
\end{equation*}
where $\Phi$ is the Newtonian potential.

The simplest condition that guarantees that $\vec\omega$ depends only on time is due to Trautmann \cite{tr}, and it reads
\begin{equation}
R^i{}^k_\bullet{}{}_{lm}=0 \label{Rcond1}
\end{equation}
or equivalently
\begin{equation}
t_{,i[}R^j{}_{kl]m}=0\ . \label{Rcond2}
\end{equation}
This equation says that parallel transport of spacelike vectors $V^i$, which are characterized by $t_{,i}V^i=0$, is path independent, a well known property of Newton's theory. If one restricts the solutions of the frame theory to spatially asymptotically flat ones (i.e. isolated systems),
then for $\lambda=0$, the conditions \eqref{Rcond1}-\eqref{Rcond2} are automatically satisfied, and Newtonian gravity is obtained.

What is the physical meaning of $\lambda$ ? One interpretation is to consider all quantities having physical dimensions. Then $\lambda$ can be identified with $1/c^2$. To avoid confusions about the meaning of $c \to\infty$, it is preferable to express the freedom of choosing units as certain similarity transformations as discussed in \cite{Ehl97}. With this in mind, the  Newtonian limit of a family
\begin{equation*}
\{g^{ik}(\lambda,x^j),T^{ik}(\lambda,x^j)\ ,\lambda>0\}
\end{equation*}
of solutions of Einstein's theory can be defined provided that $g^{ik}(\lambda,x^j)$, $T^{ik}(\lambda,x^j)$ and $t_{ik}=-\lambda g_{ik}(\lambda,x^j)$ have, together with a number of derivatives, a limit for $\lambda=0$, and these limits satisfy the axioms (i)-(v) above for frame theory.
Furthermore, if the family is differentiable in $\lambda$, then one can obtain post-Newtonian expansions.

\section{The Newtonian limit}

In this section, we describe the approach taken in \cite{Oli06,Oli07,Oli08} to mathematically justify the Newtonian limit
for isolated systems.
The articles \cite{Oli06,Oli07,Oli08} cover the situation where the matter model is a perfect fluid, and so, this is what we restrict
ourselves to in the following sections. However, it should be noted that the results of \cite{Oli06,Oli07,Oli08}
do not rely critically on the matter model, and can be extended to any other matter model that is well-posed,
admits solutions with compact support, and has a matter propagation speed that remains bounded as $c \rightarrow \infty$.
Vlasov matter is an example of a model that satisfies these conditions.

As shown in \cite{Oli06}, the Einstein-Euler equations, which govern a gravitating perfect fluid, can be written in the
following rescaled form
\leqn{EEeqn}{ G^{ij} = 2\ep^4 T^{ij} \AND
\nabla_{i} T^{ij} = 0, } where \eqn{EEdefs}{ T^{ij} = (\rho + \ep^2 p)v^i v^j
+ p g^{ij} \AND v^i v_i = -\frac{1}{\ep^2}. }
In this formulation, the fluid four-velocity
$v^i$, the fluid density $\rho$, the fluid pressure $p$, the
metric $g_{ij}$, and the coordinates $(x^i)$ $(i=1,\ldots,4)$ are
dimensionless.  Spacetime $M$ is taken to be diffeomorphic to $\Rbb^3\times [0,T)$ which is appropriate for
modeling an isolated system. The coordinates $(x^i)$ are global Cartesian
coordinates on $M\cong \Rbb^3\times [0,T)$ with the $(x^I)$
$(I=1,2,3)$ defining spatial coordinates on $ \Rbb^3$ while
$t=x^4$ is a \emph{Newtonian time coordinate} on
$[0,T)$.

As discussed in the introduction, the study of the equations \eqref{EEeqn} in the limit $\ep \searrow 0$
is called the \emph{Newtonian limit}. The overwhelming majority
of investigations of the Newtonian limit, for example see \cite{Dau64,Ehl86,Kun72,Kun76} and
reference cited therein,
have relied on formal
expansion in the parameter $\ep$. The main goal of the formal expansion
is to identify suitable gravitational-matter variables and a gauge
so that the gravitational-matter variables are well defined
in the limit $\ep \searrow 0$, and the Einstein-Euler equation \eqref{EEeqn} reduce
to the Poisson-Euler equations
\lalign{newtB}{
\del_t \rho + \del_I(\rho w^I) & = 0 \, , && (I,J=1,2,3)\label{newtB.1}\\
\rho(\del_t w^J + w^I\del_I w^J) & =
-(\rho\del^J\Phi + \del^J p) \, , && (\del^I = \delta^{IJ}\del_J )\label{newtB.2} \\
\Delta \Phi &=   \rho \, , && (\Delta = \del_I \del^I) \label{newtB.3} \,
 }
of Newtonian gravity in the limit $\ep \searrow 0$. However, as remarked in the introduction,
this is far from what one actually wants,
which is to be assured of the existence of a sufficiently large set of one-parameter families of
solutions depending on $\ep$ to the Einstein-Euler equations that exists on a common piece of
spacetime of the form $M\cong  \Rbb^3\times [0,T)$, and  converge, in a suitable sense, as $\ep \searrow 0$
to a solution of the Poisson-Euler equations on $M$.
The sufficiently large set of one-parameter families of solutions should be large enough so that
there are not any unreasonable restrictions on the choice of the initial data for the matter, which in
the perfect fluid case is the fluid density and spatial three-velocity. If this can be achieved, then this is what
is meant by a \emph{rigorous Newtonian limit}.

The rigorous analysis of the Newtonian limit starts by viewing the Einstein-Euler equations
as an initial value problem for a gauge hyperbolic system.
The main difficulty in obtaining a rigorous Newtonian limit is that the limit $\ep \searrow 0$
is a singular limit for the Einstein-Euler equations. Roughly speaking, this means
that terms of the form $1/\ep$ will appear in the evolution equations, and these terms
become unbounded as $\ep \searrow 0$. This leads to the following mathematical problems
that must be sequentially solved.
\begin{itemize}
\item[(i)]
The first problem is to find
a one parameter family of initial data
\leqn{probi1}{
\{ g_{ij}^\ep|_{t=0},\del_t g_{ij}^{\ep}|_{t=0},\rho_\ep|_{t=0},v^i_\ep|_{t=0}\}
\qquad 0<\ep < \ep_0
}
on the spacelike hypersurface
\eqn{probi2}{
\Sigma_0 = \{\, (x^I,0)\, |\, (x^I)\in \Rbb^3\, \}
}
that solves the constraint equations, and converge, in an appropriate sense, as $\ep \searrow 0$ to
Newtonian initial data.
\item[(ii)] Once problem (i) is solved and an appropriate gauge is chosen,
local existence theorems for hyperbolic partial differential equations
show that the one parameter family of initial data \eqref{probi1} generates a one-parameter family of solutions
\leqn{probii1}{
\{ g_{ij}^\ep (x,t),\rho_\ep(x,t),v^i_\ep(x,t)\} \qquad 0<\ep<\ep_0
}
on a spacetime region of the form
\eqn{probii2}{
(x,t) \in M_\ep = \Rbb^3 \times [0,T_\ep).
}
The second problem is to show that there exists a $T_*>0$ such that $T_\ep \geq T_*$ for all $0<\ep <\ep_0$, because otherwise
the one parameter family of solutions \eqref{probii1} would not exist on a common piece of spacetime, and consequently, it
would be impossible to even discuss the limit $\ep \searrow 0$.

\item[(iii)] A solution to problem (ii) guarantees the existence of a common piece of spacetime
$M_* = [0,T_*)\times \Rbb^3$ on which
the one-parameter family of solutions \eqref{probii1} to the Einstein-Euler equations exist. So once problem (ii)
is solved, the final problem
is to demonstrate the convergence on $M_*$ as $\ep \searrow 0$ of the Einstein-Euler solutions \eqref{probii1} to solutions of the Poisson-Euler
equations. This convergence should be uniform in $\ep$, and thus, part of the problem is
 to show the existence of an order $\ep$ error
estimate that measure the difference between the fully relativistic and Newtonian solutions.
\end{itemize}
A positive resolution to these three questions is the minimum requirement for a rigorous Newtonian limit.
More information about the limit $\ep \searrow 0$ would certainly be desirable, and, in some cases,
essential for answering questions about the behaviour of the Newtonian limit. For instance,
rather than a lower bound $T_*$, it would be highly desirable to determine the maximal interval $T(\ep) \geq T_*$ on
which the family of solutions \eqref{probii1} exist, and converge to solutions of the Poisson-Euler equations.
However, this is an extremely difficult problem that is at least as difficult as determining the long time
behavior of the Poisson-Euler equations, which is still an unresolved problem.

There have been various methods proposed to solve the three problems listed above. Each method is dependent on
the gauge and the gravitational-matter variables that are chosen. As stated above, we will outline the method used in
\cite{Oli06}.
For different approaches, see \cite{Ren94,ILR}. The analysis in \cite{Oli06} starts by replacing
the metric $g_{ij}$ and the fluid velocity
$v^i$ with new variables that are compatible with the limit
$\ep \searrow 0$. The new gravitational variable is a density $\ufb^{ij}$
defined via the formula
\leqn{metrecA}{
g^{ij} = \frac{\ep}{\sqrt{-\det(Q)}}Q^{ij}
}
where
\leqn{metrecB}{
Q^{ij} = \begin{pmatrix} \delta^{IJ} & 0 \\ 0 & 0 \end{pmatrix}
+  \ep^2 \begin{pmatrix} 4 \ufb^{IJ} & 0 \\ 0 & -1 \end{pmatrix}
+ 4\ep^3 \begin{pmatrix} 0 & \ufb^{I4} \\ \ufb^{J4} & 0 \end{pmatrix}
+ 4 \ep^4\begin{pmatrix} 0 & 0 \\ 0 & \ufb^{44} \end{pmatrix} .
}
From this, it not difficult to see that the density $\ufb^{ij}$ is equivalent to the metric $g_{ij}$
for $\ep > 0$, and is well defined at $\ep =0$.
For the fluid,  a new velocity variable $w^i$ is defined by
\leqn{wdef.intro}{
v^I = w^I \AND w^4=\frac{v^4-1}{\ep}\, .
}

The choice of the variables $\ufb^{ij}$ for the gravitational field was inspired by the gravitational variables
discovered by J\"{u}rgen, see Figure \ref{JEnotes} at the end of the article. To see the relationship, we let $\ufb_e^{ij}$ denote J\"urgen's variables.
In terms of these variables, the metric is given by the formula
\leqn{EhlersvarsA}{
g^{ij} =  \frac{\ep}{\sqrt{-\det(P)}}P^{ij},
}
where
\leqn{EhlersvarsB}{
P^{ij} = \begin{pmatrix} \delta^{IJ}  & 0 \\
0 & 0 \end{pmatrix}
+  \ep^2 \begin{pmatrix} 0 & 0 \\ 0 & -1 \end{pmatrix}
+ 4 \ep^4 \begin{pmatrix} \ufb_e^{IJ} & \ufb_e^{J0} \\ \ufb_e^{J0} & \ufb_e^{00}
\end{pmatrix}.
}
The formulas \eqref{metrecA}, \eqref{metrecB}, \eqref{EhlersvarsA}, and \eqref{EhlersvarsB} show
that $\ufb^{ij}$ are related to the $\ufb_e^{ij}$ through a simple rescaling of the
components by an appropriate power of $\ep$.

For technical reasons, an isentropic equation of
state
 \leqn{eos}{ p =
K\rho^{(n+1)/n}, } where $K \in \Rbb_{>0}$ and $n\in \Nbb$ is assumed. This
allows for the use of
a technique of Makino \cite{Mak} to regularize the fluid
equations through the use of the following fluid density variable
\leqn{dendef}{ \rho = \frac{1}{\bigl(4Kn(n+1)\bigr)^n}\alpha^{2n}. }
This choice makes it possible
to formulate the Euler equations as a hyperbolic system
that is regular across the
fluid-vacuum interface.
In this way,
it is possible to construct solutions to the Einstein-Euler
equations that represent compact gravitating fluid bodies (i.e. stars)
both in the Newtonian and
relativistic setting \cite{Mak,Ren92}. In the Newtonian setting,
this is straightforward to see. Using \eqref{eos}
and \eqref{dendef}, the Poisson-Euler equations \eqref{newtB.1}-\eqref{newtB.3} become
\lalign{newtA}{
\del_t\alpha &= -w^I\del_I
\alpha -\frac{\alphah}{2n}\del_I w^I, \label{newtA.1} \\
\del_t w^J &= -\frac{\alpha}{2n}\del^J\alpha - w^I\del_I
w^J -\del^J\Phi \label{newtA.2}, \\
\Delta \Phi &=  \rho, \label{newtA.3} }
which is readily seen to be regular even across regions where $\alpha$
vanishes.

It is important to note that the Makino technique only produces
a particular subclass of solutions to the Euler equations representing
compact fluid bodies. In particular, the type of fluid solutions
obtained by this method have freely falling boundaries \cite{Ren92}, and hence,
do not include static stars of finite radius. On the one hand, this makes the Makino
technique far from ideal, but on the other,  it is the only
method presently available for generating relativistic solutions that represent
gravitating, compact
fluid bodies. However, in trying to understand the
Newtonian limit and post-Newtonian approximations, these solutions
are almost certainly general enough to obtain a comprehensive
understanding of the mathematical issues involved in the Newtonian
limit and post-Newtonian approximations. This is because the
mathematical difficulties in studying the post-Newtonian expansions do not
rely heavily on the structure of the matter model. As stated above, the matter model
need only be
well-posed, admit solutions with compact support, and have a matter propagation
speed that remains bounded in the limit $\ep \searrow 0$.

The approach taken in \cite{Oli06,Oli07,Oli08} relies on a particular nonlocal first order formulation for
the Einstein-Euler system. This formulation starts by the introduction of the following vector
\eqn{fovars}{
V = ( \ep \del_t \ufb^{ij}, \del_I \ufb^{ij}, \ep \ufb^{ij},\alpha,w^i),
}
which contains both the gravitational and matter variables. Assuming a harmonic gauge,
a straightforward calculation shows that
$V$ satisfies an equation of the form
\leqn{Veqn1}{
b^0(\ep V)\del_t V = \frac{1}{\ep}c^I \del_I V + b^{I}(\ep,V)\del_I V + f(\ep,V) + \frac{1}{\ep}g(\rho)
}
where $b^0(\ep V)$ and $b^I(\ep,V)$ are symmetric matrices that depend analytically on $\ep V$ and $(\ep,V)$, respectively,
$f$ is a map
that depends analytically on $(\ep,V)$,
\alin{Veqn2}{
c^I &= \begin{pmatrix} \delta^{IJ} & 0          & 0 & 0 & 0  \\
                         0        & \delta^{IJ}& 0 & 0 & 0  \\
                         0        & 0          & 0 & 0 & 0  \\
                         0        & 0          & 0 & 0 & 0  \\
                         0        & 0          & 0 & 0 & 0
\end{pmatrix},
\intertext{and}
g &= \bigl(-\delta^i_4\delta^j_4\rho,0,0,0,0\bigr)^T.
}
This formulation has the virtue that the are only two terms which contain $1/\ep$, and these
terms are of a particularly simple structure. However,
from an analysis point of view, the presence of
the term
$\frac{1}{\ep} g(\rho)$ is problematic. Fortunately, this term can be
removed
by subtracting off the gravitational contribution from
the Newtonian potential in the following fashion
\leqn{Weqn1}{
W = ( \ep \del_t \ufb^{ij}, \del_I \ufb^{ij}-\delta^i_4\delta^j_4\del_I\Phi, \ep \ufb^{ij},\alpha,w^i)
}
where $\Phi$ solves
\eqn{Weqn2}{
\Delta\Phi = \rho.
}
This effectively moves the problem term $\frac{1}{\ep}g(\rho)$ into the $c^I\del_I W$ term.
In terms of the new variable $W$, the Einstein-Euler equations become
\leqn{Weqn3}{
b^0(\ep W)\del_t W = \frac{1}{\ep}c^I \del_I W + b^{I}(\ep,W)\del_I W_\ep + F(\ep,W),
}
where $F$ is a nonlocal map that depends analytically on $(\ep,W)$.
The importance of this nonlocal symmetric hyperbolic formulation of the Einstein-Euler equations
is that it is possible, given suitable initial data, to use energy estimates to establish the existence
of solutions on a time interval that is independent of $\ep$. In fact, equation
\eqref{Weqn3} is a particular example of a class of singular hyperbolic equations
for which the limit $\ep \searrow 0$ has been well studied \cite{KM82,Kreiss80,Scho86,Scho88,Scho94}.
However, it is important to note the results of these articles
do not directly apply equation \eqref{Weqn3}. The reason for this is that the initial data for the system
\eqref{Weqn3} must include a $1/r$ piece for the metric, and hence, cannot lie in the Sobolev space $H^k$ on
which the results of \cite{KM82,Kreiss80,Scho86,Scho88,Scho94} are based.

As discussed above, the first problem is to generate suitable initial data for the evolution equation \eqref{Weqn3}.
The method used in \cite{Oli06,Oli08} for generating this data is based on
a method introduced by Lottermoser \cite{Lott}. The result is a positive solution to problem (i) listed above.
We will not discuss the details here and only outline the main result.

First, let $\{\delta,s,R,k\}$ be a fixed set of numbers satisfying
$-1 < \delta < -1/2$, $s\in \mathbb{Z}_{\geq 2}$, $R>0$,
$k\in \mathbb{Z}_{\geq 3+s}$. Then it is shown in \cite{Oli08}, that
for any choice of the \emph{free initial data}
\leqn{idata1}{
\bigl(\mathfrak{z}^{IJ},\mathfrak{z}^{IJ}_4 \in H^k_{\delta-1},
\alpha_0,w_0^I\bigr) \in H^{k+1}_\delta\times H^k_{\delta-1} \times
H^k_{\delta-1}\times H^k_{\delta-1}
}
where the density has compact support $\supp\, \alpha_0 \subset B_R$,
there exists a one parameter family of determined data
\leqn{idata2}{
\bigl(\phi_{\ep},\wf^K_\ep,w^4_{0,\ep}\bigr) \in H^{k+1}_\delta\times H^{k+1}_\delta\times H^{k}_{\delta-1}
\quad \ep \in (-\ep_0,\ep_0)
}
that depends analytically on $\ep$. The gravitational variable $\ufb^{ij}_\ep$ and its first time derivative are determined
on the hypersurface $\Sigma_0$ in terms
of the free and determined initial data as follows:
\lalign{idata3}{ (\ufb^{ij}_\ep) &
=\begin{pmatrix} \zf^{IJ} &
\wf^I_\ep \\
\wf^J_\ep &
\phi_\ep
\end{pmatrix}, \label{idata3.1}
\intertext{and}
(\ep \del_t\ufb^{ij}_\ep) & = \begin{pmatrix} \zf^{IJ}_4 & -
\del_{K}\zf^{KI}
\\ -\del_{K}\zf^{KJ}& -\del_K\wf^K_\ep
\end{pmatrix}. \label{idata3.2}
}
Together, \eqref{idata3.1}, \eqref{idata3.2}, and
\leqn{idata4}{
(\rho_0,w^I_0,w^4_{0,\ep})
}
solve the gravitational constraint
equations $G^{I0}-2\ep^4 T^{I0}=0$, the harmonic condition $g^{ij}\Gamma_{ij}^k=0$,
and the fluid velocity normalization $v^i v_i = -1/\ep^2$ on $\Sigma_0$ for $-\ep_0 < \ep  < \ep_0$.
Moreover, $\phi_\ep$, $\wf^I_\ep$, and $w^4_{0,\ep}$ satisfy the equations \eqn{idatA3a}{
\Delta\phi_0 = \rho_0 + \del^2_{IJ}\zf^{IJ}\, , \quad \Delta\wf^I_0 = -\del_L\zf^{LJ}_4
, \AND w^4_{0,0} =  0 \, , } respectively, at $\ep=0$. We also note that the free data \eqref{idata1} can
be chosen so that any one of the $\mathfrak{z}^{IJ}$, $\mathfrak{z}^{IJ}_4$,
$\alpha_0$, and $w_0^I$ depends analytically on $\ep$. Indeed, as will be discussed below,
the rigorous post-Newtonian expansions require that the gravitational
free initial data $\mathfrak{z}^{IJ}$, $\mathfrak{z}^{IJ}_4$ is $\ep$-dependent.

With the question of generating suitable initial data resolved, the next step is to resolve
problem (ii) which concerns finding a lower bound on the time of existence for solutions
$W_\ep(x,t)$
to \eqref{Weqn3} generated from the $\ep$-dependent initial data \eqref{idata3.1}-\eqref{idata4}.
This requires decomposing $W_\ep$ into the sum
of a time independent $1/r$ term and a time varying $1/r^2$ term:
\eqn{Weqn4}{
W_\ep(x,t) = W_{-1,\ep}(x) + W_{-2,\ep}(x,t),
}
where
\eqn{Weqn5}{
W_{-1}(x) = (0,0,\ep\ub^{ij}_\ep|_{t=0},0,0) \AND W_{-2,\ep}(x,t) = W_\ep(x,t)-W_{-1}(x).
}
The initial data given by \eqref{idata3.1}-\eqref{idata4} then implies that
\eqn{Weqn6}{
\norm{W_{-1,\ep}}_{H^{k+1}_\delta} \lesssim 1 \AND \norm{W_{-2,\ep}(0)}_{H^k_{\delta-1}} \lesssim 1 \quad \forall \ep \in [0,\ep_0).
}
This bound on the initial data and the special form of equation \eqref{Weqn3}
are the key to deriving $\ep$-independent energy estimates. These energy estimates
provide a uniform bound on the solution for some time $T_*$ which may, a priori, be
larger than the time of existence. But, by continuation principle for symmetric hyperbolic
equations, solutions can be continued as long as they remain bounded in the $W^{1,\infty}$ norm, and
so, the $\ep$-independent energy estimates together with the Sobolev inequality guarantee the existence of a classical
solution
\eqn{Weqn7}{
W_\ep(x,t) \in C^1(\Rbb^3\times [0,T_*) )  \quad 0 < \ep <\ep_0
}
to \eqref{Weqn3} that satisfies the uniform bound
\eqn{Weqn6a}{
\norm{\ufb^{ij}_\ep(t)}_{L^6} + \norm{W_{-2,\ep}(t)}_{H^k}
+\ep \norm{\del_t W_\ep(t)}_{H^{k-1}} \lesssim 1 \quad \forall \ep \in (0,\ep_0).
}
This not only resolves problem (ii) stated above, but also provides uniform estimates
 for the one parameter family of solutions to the Einstein-Euler equations determined by
$W_\ep(x,t)$.

For initial data that is general as \eqref{idata3.1}-\eqref{idata4}, energy estimates alone are not
sufficient to resolve problem (iii), and establish convergence to a Newtonian solution.  To obtain
convergence, the energy estimates
must be combined with dispersive estimates for the wave equation \cite{Oli08}. Together, these
estimates can be used to establish the existence of
a solution $\{\rhot(x,t),\wt^I(x,t),\Phit(x,t)\}$
to the Poisson-Euler equations
\alin{newtC}{
\del_t \rhot + \del_I(\rho \wt^I) & = 0 \, , \\
\rhot(\del_t \wt^J + \wt^I\del_I \wt^J) & =
-(\rhot \del^J\Phit + \del^J p) \, ,  \\
\Delta \Phit &=   \rhot \,  \,
 }
with initial conditions
\eqn{newtCidata}{
\rhot|_{t=0} = (4Kn(n+1))^{-1}\alpha_0^{2n}, \quad \wt^I|_{t=0} =w_0^I
}
and a solution $\ufbt_\ep(x,t)$ of the wave equation
\leqn{mthb}{
\ep^2\del_t^2 \ufbt_\ep^{ij} - \Delta \ufbt_\ep^{ij} = -\delta^i_4\delta^j_4 \rhot +\ep^2\delta^i_4\delta^j_4\del_t^2\Phit,
}
with initial conditions
\alin{mth3b}{
\ufbt^{ij}_\ep\bigl|_{t=0} & = \delta^i_I\delta^j_J\zf^{IJ}
-2\Delta^{-1}\del_I\zf_4^{IJ}\delta^{(i}_4\delta^{j)}_J + \delta^{i}_4\delta^j_4
(\Phit\bigl|_{t=0}+\Delta^{-1}\del^2_{IJ}\zf^{IJ}), \\
\del_t\ufbt^{ij}_\ep\bigl|_{t=0} & = \frac{1}{\ep} \bigl(\delta^i_I\delta^j_J\zf_4^{IJ}
-2\del_I\zf^{IJ}\delta^{(i}_4\delta^{j)}_J + \delta^{i}_4\delta^j_4
\Delta^{-1} \del^2_{IJ}\zf_4^{IJ} \bigr) + \delta^i_4\delta^j_4\del_t\Phit\bigl|_{t=0} ,
}
such that the following uniform estimates hold
\leqn{mth4}{\norm{\ufb^{ij}_\ep(t)-\ufbt^{ij}_{\ep}}_{L^6}+
\norm{\del_I\ufb_\ep(t)- \del_I\ufbt^{ij}_{\ep}(t)}_{H^{k-2}}
+ \norm{\ep\del_t\ufb^{ij}_\ep(t)- \ep\del_t\ufbt^{ij}_\ep(t)}_{H^{k-2}}
\lesssim \ep }
and
\leqn{mth5}{\norm{\rho_\ep(t)-\rhot(t)}_{H^{k-2}} +
\norm{w^I_\ep(t)-\wt^I(t)}_{H^{k-2}} + \norm{w^4_\ep(t)}_{H^{k-2}} \lesssim \ep,
}
for all $(t,\ep) \in [0,T_*)\times (0,\ep_0)$.  It is this last estimate \eqref{mth4} that
resolves problem (iii) and established rigorously the convergence as $\ep \searrow 0$
of the fully relativistic solution of the Einstein-Euler equations determined by
$W_\ep(x,t)$ to a solution of the Poisson-Euler equations. Clearly, it also provides
an order $\ep$ error estimate that measures the difference between the relativistic
and Newtonian solutions. We also note that
\leqn{mth5a}{
\norm{\ufbt^{ij}_{\ep}}_{L^6} + \norm{\del_I\ufbt^{ij}_{\ep}(t)}_{H^{k-2}}
\lesssim 1
}
for $t\in [0,T_0)$. Combined, the formula \eqref{metrecA}, and the  estimates \eqref{mth4} and \eqref{mth5a}  show
that
\leqn{mth5b}{
\norm{g_\ep^{ij}(t)-\delta^{iI}\delta^{jJ}\delta_{IJ}}_{L^6} + \norm{\del_I g_\ep^{ij}(t)}_{H^{k-2}} \lesssim \ep
}
and
\leqn{mth5c}{
\norm{\ep^2 g^\ep_{ij}(t)+\delta^4_i\delta^4_j}_{L^6} + \norm{\ep^2 \del_I g^\ep_{ij}(t)}_{H^{k-2}} \lesssim \ep
}
for all $t\in [0,T_0)$. It also follows from the estimates \eqref{mth4}-\eqref{mth5a}, and the fact the support of $\rho_e$
stay uniformly bounded, that the stress energy tensor $T^{ij}_\ep$ satisfies
\leqn{math5d}{
\norm{T^{ij}_\ep(t)- \delta^i_I\delta^j_J(\rhot(t) \wt^I(t) \wt^J(t) + p(t)\delta^{IJ})}_{H^{k-1}} \lesssim \ep
}
for all $t\in [T_0,\ep)$.

The estimate \eqref{mth4} is not strong enough to guarantee that the gravitational
variables $\ufb^{ij}$ converge to a particular limit. It only says the gravitational part of the
fully relativistic solution is approximated to order $\ep$ by a solution of
the singular wave equation \eqref{mthb}. However, after an arbitrarily small
time, the dispersive effects of the wave equation are strong enough to
allow for convergence, at least for the spatial and temporal derivatives.
This can be seen by the following estimate: given any $t_0\in (0,T_*)$,
the solution $\ufbt^{ij}$ satisfies
\leqn{fcov1}{
\norm{\del_I\ufbt^{ij}(t)-\delta^i_4\delta^j_4 \del_I\Phi(t)}_{W^{k-2,\infty} } + \ep \norm{\del_t\ufbt^{ij}(t)-\delta^i_4\delta^j_4 \del_t \Phit(t)}_{W^{k-2,\infty}}
\lesssim_{t_0} \ep
}
for all $t\in (t_0,T_*)$. Moreover, restricting to any ball $B_\Lambda \subset \Rbb^3$ of radius $\Lambda$, the above estimate improves to
\leqn{fcov2}{
\norm{\del_I\ufbt^{ij}(t)-\delta^i_4\delta^j_4 \del_I\Phi(t)}_{W^{k-2,\infty} } \lesssim_{t_0,\Lambda} \ep^{3/2},
}
and
\leqn{fcov3}{
\norm{\del_t\ufbt^{ij}(t)-\delta^i_4\delta^j_4 \del_t \Phit(t)}_{W^{k-2,\infty}(B_\Lambda)}
\lesssim_{t_0,\Lambda} \sqrt{\ep}
}
for all $t \in (t_0,T_*)$. We also note that in the estimates \eqref{mth4}-\eqref{mth5}, and \eqref{fcov1}-\eqref{fcov3},
each time a time derivative
is taken, the estimate gets worse by a power of $\ep$. For example, it follows from \eqref{fcov3}
that
\leqn{fcov4}{
\norm{\del_t^2\ufbt^{ij}(t)-\delta^i_4\delta^j_4 \del_t^2 \Phit(t)}_{W^{k-3,\infty}(B_\Lambda)}
\lesssim_{t_0,\Lambda} \frac{1}{\sqrt{\ep}}
}
for all $t \in (t_0,T_*)$

Together, the estimates \eqref{mth4}, and \eqref{fcov1}-\eqref{fcov3} establish
that $\del_I \ub^{ij}_\ep$ and $\del_t \ub^{ij}_\ep$ converge to $\delta_4^i\delta_4^j \del_I \Phit$ and
$\delta_4^i\delta_4^j\del_t \Phit$, respectively, as $\ep \searrow 0$, at least after waiting a short time.
They also show that the Christoffel symbols satisfy
\leqn{christ1}{
\norm{\Gamma_{ij}^k -\delta_i^4 \delta_j^4 \delta^{Ik} \del_I \Phit(t)}_{W^{k-3,\infty}} \lesssim_{t_0} \ep
}
and
\leqn{christ2}{
\norm{\del_t\Gamma_{ij}^k - \delta_i^4\delta_j^4 \delta^{Ik} \del_I \del_t\Phit(t)}_{W^{k-4,\infty}(B_\Lambda)}
\lesssim_{t_0,\Lambda} \sqrt{\ep}
}
for all $t \in (t_0,T_*)$. In particular, this implies that on the spacetime region $M_{\Lambda,t_0} = B_\Lambda \times (t_0,T_*)$,
both the Christoffel symbols and the curvature components converge as $\ep \searrow 0$, and hence this limit
satisfies the assumptions on J\"urgen's frame theory on the bounded spacetime region $M_{\Lambda,t_0}$.

It is possible to prove convergence of the relativistic solution to a Newtonian
solution using energy estimates alone if the initial data is suitably restricted.
Indeed, as shown in \cite{Oli06}, if the free initial data is
chosen so that
\eqn{free}{
\zf^{IJ} = \text{O}(\ep) \AND \zf^{IJ}_4 = \text{O}(\ep)\quad \text{as $\ep\searrow 0$},
}
then $W_\ep(t,x)$ satisfies
\leqn{Wt1}{
\norm{\del_t W_\ep|_{t=0}}_{H^{k-1}_\delta} \lesssim 1 \quad \ep \in (0,\ep_0).
}
This uniform estimate combined with energy estimates can be used
to improve
\eqref{mth4} and \eqref{mth5} to
\leqn{mth6}{\norm{\rho_\ep(t)-\rhot(t)}_{H^{k-1}} +
\norm{w^I_\ep(t)-\wt^I(t)}_{H^{k-1}} + \norm{w^4_\ep(t)}_{H^{k-1}} \lesssim \ep}
and
\leqn{mth7}{\norm{\ufb^{ij}_\ep(t)-\delta^i_4\delta^j_4\Phit(t)}_{L^6}+
\norm{\del_I\ufb^{ij}_\ep(t)- \delta^i_4\delta^j_4\del_I\Phit(t)}_{H^{k-1}} + \norm{\ep \del_t \ufb^{ij}_\ep }_{H^{k-1}}
\lesssim \ep. }
In particular, this shows that the condition \eqref{Wt1} on the initial data
implies that both the gravitational $\ufb^{ij}_\ep$ and the matter
variables $\{\rho_\ep, w^i_\ep\}$ converge as $\ep \searrow 0$. We also note that with this restricted data,
the estimates \eqref{christ1} and \eqref{christ2} improve to
\eqn{christ3}{
\norm{\Gamma_{ij}^k -\delta_i^4\delta_j^4 \delta^{Ik} \del_I \Phit(t)}_{H^{k-1}} \lesssim \ep
}
for all $t\in [0,T_*)$ and
\eqn{christ4}{
\norm{\del_t\Gamma_{ij}^k - \delta_i^4\delta_j^4 \delta^{Ik} \del_I \del_t\Phit(t)}_{W^{k-3,\infty}}
\lesssim_{t_0,\Lambda} \sqrt{\ep}
}
for all $t\in (t_0,T_*)$, respectively. This gives convergence on the space time region
$M_{t_0} = \Rbb^3 \times (t_0,T_*)$.

To get convergence, in the sense of frame theory, on
the whole spacetime slab $M_*=\Rbb^3\times [0,T_*)$, requires that initial data is chosen
so that both the first and second time derivatives of $W_\ep$ at $t=0$
are bounded as $\ep \searrow 0$. This type of additional restriction of the time derivatives
is also what is needed to obtain rigorous post-Newtonian expansions, which will be discussed
in the next section.

\section{Post-Newtonian expansions}

That the boundedness condition \eqref{Wt1} on the first
time derivative of $W_\ep$ at $t=0$ implied the existence of
a $0^{\text{th}}$ order expansions in $\ep$ is a manifestation of
\emph{Kreiss's bounded derivative principle} \cite{BK,Kreiss91}. This principle
states that using energy estimates alone, the problem of generating $\ell^{\text{th}}$ order expansions
in $\ep$  for solutions to an equation of the form \eqref{Weqn3} can be reduced
to the problem of finding initial data that satisfies
\leqn{dWl}{
\norm{\del_t^p W_\ep|_{t=0}}_{H^{k-p}_{\delta-1}} \lesssim 1
}
for $p=1,2,\ldots,\ell+1$.
This process of
choosing initial data to satisfy \eqref{dWl} is called \emph{initialization}.
This shows that the existence problem for the post-Newtonian expansions can be
replaced by the problem of finding initial data that can be properly initialized.
We note that the conditions \eqref{dWl} effectively amounts to additional elliptic equations that the initial data
must satisfy in addition to the usual constraint equations. For an example of these additional elliptic
equations on the initial data, see equation (6.4) in \cite{Oli07}.

One of the results of \cite{Oli07} is to show that it is possible to construct
initial data that satisfies
\leqn{dW3}{
\norm{ \del_t^p W_\ep|_{t=0} }_{H^{k-p}_{\delta-1}} \lesssim 1 \quad p=0,1,2,3
}
By Kreiss's bounded derivative principle, this is enough to prove the existence of a
first post-Newtonian expansion for the Einstein-Euler
equations. More specifically, it is possible to show that the solution
\eqn{EEsol}{
\{\ufb^{ij}_\ep(x,t),\alpha_\ep(x,t),w^i_\ep(x,t)\} \qquad (x,t)\in M_*
}
admit  convergent expansions (uniform for $0 < \ep \leq \ep_0$)  of the form
\lalign{pnexp}{
\ufb^{ij}_\ep&= \delta^i_4\delta^j_4\Phih + \sum_{q=1}^2
\ep^q \oset{q}{\ufb}{}^{ij} +
\sum_{q=3}^\infty \ep^q \oset{q}{\ufb}{}_\ep^{ij},  \label{pnexp.1}  \\
\ep^\nu\del_t^\nu\del_I\ufb^{ij}_\ep &= \ep^\nu
\delta^i_4\delta^j_4\del_t^\nu\del_I\Phih + \sum_{q=1}^2
\ep^{q+\nu} \del_t^\nu\del_I\oset{q}{\ufb}{}^{ij} +
\sum_{q=3}^\infty \ep^{q+\nu} \del_t^\nu\del_I\oset{q}{\ufb}{}_\ep^{ij}
&& \nu=0,1,   \label{pnexp.2} \\
\ep^\nu \del_t^\nu \ufb^{ij}_\ep &= \ep^\nu \delta^i_4\delta^j_4
\del_t^\nu \Phih + \sum_{q=1}^2
\ep^{q+\nu} \del_t^\nu \oset{q}{\ufb}{}^{ij} +
\sum_{q=3}^\infty \ep^{q+\nu} \del_t^\nu \oset{q}{\ufb}{}_\ep^{ij}
&& \nu=1,2, \label{pnexp.3}
}
and
\lalign{pnexpa}{
\del_t^\nu \alpha_\ep & =
\del_t^\nu \alphah + \sum_{q=1}^2 \ep^q\del_t^\nu \oset{q}{\alpha}
+\sum_{q=3}^\infty \ep^q\del_t^\nu \oset{q}{\alpha}_\ep &&
\nu = 0,1,  \label{pnexp.4} \\
\del_t^\nu w^i_\ep & =
\del_t^\nu \wh{}^i + \sum_{q=1}^2 \ep^q\del_t^\nu \oset{q}{w}{}^i
+\sum_{q=3}^\infty \ep^q\del_t^\nu \oset{q}{w}{}_\ep^i &&
\nu = 0,1,  \label{pnexp.5}
}
where the first expansion is convergent in $C^0([0,T_*);L^2_\delta)$,
and the rest are convergent in $C^0([0,T_*);H^{k-4})$. Here, the expansions coefficient $(q=1,2)$
$\{\oset{q}{\ufb}{}^{ij}(t),\oset{q}{\alpha}(t),\oset{q}{w}{}^i(t)\}$
satisfies a linear (nonlocal) symmetric hyperbolic system
that only depends on $\{\alphat(t),\wt{}^I(t),\Phit(t)\}$ if $q=1$,
and $\{\alphat(t),\wt{}^I(t),\Phit(t),\oset{1}{\ufb}{}^{ij}(t),\oset{1}{\alpha}(t),\oset{1}{w}{}^i(t)\}$ if $q=2$,
while for $q\in \Zbb_{\geq 3}$,  $\{\oset{q}{\ufb}{}_\ep^{ij}(t),
\oset{q}{\alpha}_\ep(t),\oset{q}{w}{}_\ep^i(t)\}$ satisfies a linear (non-local) symmetric hyperbolic system
that only depends on $\ep$,  $\{\alphat(t),\wt{}^I(t),\Phit(t)\}$,
 $\{\oset{p}{\ufb}{}^{ij}(t),\oset{p}{\alpha}(t),\oset{p}{w}{}^i(t)\}$
for $p=1,2$, and
 $\{\oset{p}{\ufb}_\ep{}^{ij}(t),\oset{p}{\alpha}_\ep(t),
\oset{p}{w}{}_\ep^i(t)\}$ for $p=3,4,\ldots,q-1$.

Since, the $\{\oset{q}{\ufb}{}^{ij}(t),\oset{q}{\alpha}(t),\oset{q}{w}{}^i(t)\}$ are $\ep$ independent,
the expansions \eqref{pnexp.1}-\eqref{pnexp.5} do, in fact, coincide with the first post-Newtonian expansion.
This can be seen by introducing the following
$\ep$-independent quantities:
\eqn{hkdef}{
\oset{q}{h}{}^{ij} = \bigl(4\oset{q}{\ufb}{}^{ij} - 2\eta_{k\ell}\oset{q}{\ufb}{}^{k\ell}\eta^{ij}\bigr) \qquad q=1,2,
}
where $(\eta_{ij}) = \diag(1,1,1,-1)$. From these expression and the expansions
\eqref{pnexp.1}-\eqref{pnexp.5}, it follows that (see \eqref{metrecA}-\eqref{metrecB})
the metric $g_{ij}$ is given by
\alin{gexp}{
g_{44} & = -\frac{1}{\ep^2}-2\Phih - \ep \oset{1}{h}{}^{44} - \ep^2
\Bigl(3\bigl(\Phih\bigr)^2 + \oset{2}{h}{}^{44} \Bigr)  + \text{O}(\ep^3),\\
g_{4I} & = \ep^2 \oset{1}{h}{}^{4I} + \ep^3\oset{2}{h}{}^{4I} + \text{O}(\ep^4),
\intertext{and}
g_{IJ} & = \delta_{IJ} -2\ep^2\delta_{IJ}\Phih -\ep^3\oset{1}{h}{}^{IJ}
-\ep^4\Bigl( \bigl(\Phih\bigr)^2\delta_{IJ}+\oset{2}{h}{}^{IJ} \Bigr) + \text{O}(\ep^5).
}
It is important to note that higher order expansions in $\ep$ can be generated for the metric $g_{ij}$ using \eqref{pnexp.1}-\eqref{pnexp.5}.
However, these higher order terms will, in general, depend on $\ep$ in a non-analytic fashion. Consequently, they represent some sort of
generalized post-Newtonian expansion, and it is not clear without further analysis how these expansions are related to
the standard post-Newtonian ones.

To make progress beyond the first post-Newtonian expansion, the simplest thing to do would be to try and prove
the existence of initial data that satisfies \eqref{dWl} for $\ell > 3$. In the present
setup using the harmonic gauge, the condition \eqref{dWl} for $\ell=4$ leads to elliptic equations
that are not solvable within the required function spaces. On the other hand, it is  expected that
with a suitable gauge choice, it should be possible to generate
post-Newtonian expansions to at least the
$2.5$ post-Newtonian order after which there are indications that
the post-Newtonian expansions will break down. For a lucid
discussion of this phenomenon see \cite{Ren92a}.

As remarked in \cite{Ren92a}, the choice of harmonic
gauge may be the reason for not being able
to reach the $2.5$ post-Newtonian order, and there may well
exist other gauges for which the conditions \eqref{dWl} can be satisfied
for $\ell > 3$. However, even if this is the case, it would still
need to be verified that these other gauges are compatible with
the singular hyperbolic energy estimates that are guaranteed
to arise.

Kreiss's bounded derivative principle which reduces the existence problem of $\ell^{\text{th}}$ order
expansions in $\ep$ to that of finding initial data satisfying \eqref{dWl} relies on energy
estimates. In our situation, we also have dispersive wave estimates coming from the fact that
the gravitational variables $\ufb^{ij}_\ep$ satisfy a wave equation. These dispersive estimates
can be used to show that the matter variables are more regular in $\ep$ as compared to the
gravitational ones. This means that it is possible to go beyond the first post-Newtonian
expansion while only satisfying \eqref{dWl} for $\ell = 3$. This type of effect was
already established in \cite{Oli08} at the Newtonian level. This is clear from the estimates
\eqref{mth4}-\eqref{mth5} which establish the existence of a Newtonian limit without need
for the condition \eqref{Wt1}. In fact, it is possible to show using the results of
both the paper \cite{Oli07,Oli08} that there exists post-Newtonian expansions to the second
order. However, it still remains to be seen what is the optimal choice of gauge
that allows one to satisfy \eqref{dWl} to as high order as possible while still being able to
use energy and dispersive estimates to generate post-Newtonian
expansions to the highest possible order.

It is well known that Newtonian gravity generalizes to the cosmological setting \cite{RS97}. Using similar techniques
as outlined above, it is also possible to rigorously establish the existence of cosmological post-Newtonian
expansions \cite{Oli09}. In contrast to the asymptotically flat case, it is possible to construct initial
data that satisfies \eqref{dWl} for arbitrarily large $\ell$. This implies that there is no obstruction
to generating rigorous post-Newtonian expansions to arbitrary order which is somewhat surprising in light
of difficulties that are encountered at both the rigorous and formal level in trying to develop
post-Newtonian expansions on asymptotically flat spacetimes beyond the order $2.5$.  On asymptotically flat spacetimes,
the problems that occur in the higher order post-Newtonian expansions are often attributed to the reaction of gravitational
radiation with itself and matter. The analysis contained in the paper \cite{Oli09} shows that this is not the complete
story as these effects are also present in the cosmological setting but do not cause similar difficulties.

\end{document}